\documentclass[%
 reprint,
 amsmath,amssymb,
 aip,
]{revtex4-1}
\usepackage{graphicx}
\usepackage{dcolumn}
\usepackage{bm}
\usepackage{xcolor}
\begin{document}
\title{Inverse Edelstein effect of the surface states of a topological insulator}
\author{Hao Geng}
\author{Wei Luo}
\author{W.Y. Deng}
\affiliation{
National Laboratory of Solid State Microstructures and Department of Physics, Nanjing University, Nanjing 210093, China\\
}
\author{L. Sheng}
\email{shengli@nju.edu.cn}
\author{R. Shen}
\author{D. Y. Xing}%
\affiliation{
National Laboratory of Solid State Microstructures and Department of Physics, Nanjing University, Nanjing 210093, China\\
}
\affiliation{Collaborative Innovation Center of Advanced Microstructures,
Nanjing University, Nanjing 210093, China}
\begin{abstract}
The surface states of three-dimensional topological insulators posses the unique property of spin-momentum interlocking. This property gives rise to the interesting inverse Edelstein effect (IEE), in which an applied spin bias
$\mu$ is converted to a measurable
charge voltage difference $V$. We develop a semiclassical theory for the IEE
of the surface states of $\text{Bi}_2\text{Se}_3$ thin films, which is
applicable from the ballistic regime to diffusive regime. We find that
the IEE efficiency ratio $\gamma=V/\mu$ exhibits universal dependence on sample size, and
approaches $\pi/4$ in the ballistic limit and $1$ in the diffusive limit.
\end{abstract}


\maketitle
\label{1}

Spintronics has been a rapidly growing field of research in the past two decades because of its potential applications  in memory, logic, and  sensing devices, which utilize both spin and charge degrees of freedom of electrons~\cite{Wolf, Zutic, Fert, Parkin, Bader,Han,Matsunaga}. Among the major tasks in spintronics, electrical detection of spin current and spin bias remains to be challenging. One method is to use the inverse spin Hall effect (ISHE),
in which a pure spin current generates a measurable transverse charge current~\cite{hir,val,kim,Wera}.
While the ISHE has been widely employed in spintronic experiments~\cite{mos,mos1,roj,cze,roj,tser},
the electrical signal generated is usually small,
e.g., the spin Hall angle $\theta_{\mathrm{sh}}=0.08$ in Pt~\cite{Liu}.
Another method that has been attracting increasing interest is the inverse Edelstein effect (IEE)~\cite{Mahfouzi, Shen}, in which spin injection induces nonequilibrium spin polarization and in turn generates a  charge current in the longitudinal direction.  The IEE has been observed in Bi~\cite{Sanchez}, which was attributed to the Rashba spin-orbit coupling on the interface.

Topological insulators (TIs)~\cite{Hasan, Qi1} and topological Kondo insulators (TKIs)~\cite{Dzero} are a new
quantum state of matter. A three-dimensional (3D) TI has a bulk insulating gap with gapless surface states, which are protected from impurity backscattering by nontrivial bulk band topology and time-reversal symmetry.
The topological surface states posses the unique property of spin-momentum locking~\cite{Qi1, Qi2, Dzero},
which are promising for applications in spintronic devices~\cite{Yokoyama, Modak}.
In 2014, large IEE was realized in bulk insulating TIs $\text{Bi}_{1.5}\text{Sb}_{0.5}\text{Te}_{1.7}\text{Se}_{1.3}$ and Sn-doped $\text{Bi}_2\text{Te}_2\text{Se}$~\cite{Shiomi}, which was interpreted as a result of the spin-momentum locking of the topological surface states.
Recently, in another experimental work~\cite{Song}, the IEE was observed on the surface of TKI $\text{Sm} \text{B}_6$.
By using a Landauer-B\"{u}ttiker like formula, Luo $et$ $al.$
theoretically studied the IEE of the surface states in the ballistic regime, and predicted that a
spin bias polarized in the $y$ direction can generate a charge current flowing in the $x$ direction~\cite{Luo},
which is in good agreement with the experimental observation~\cite{Song}.
However, the effect of impurity scattering and sample size dependence
in the IEE are not addressed in the simplified theory~\cite{Luo}.
In this Letter, we follow the model of Luo $et$ $al.$~\cite{Luo}  and employ
a semiclassical approach~\cite{Geng} to study the IEE of the topological surface states. Our analytical theory
is applicable from ballistic to diffusive regime, and may provide useful guidance for experimental study of
the IEE in 3D TIs.



\begin{figure}
\begin{center}
\includegraphics[width=6cm]{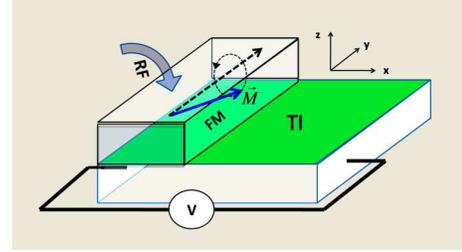}
\caption{Schematic view of the setup for observing the IEE.  A TI thin film is covered partly by a ferromagnetic metal.
  When the magnetization of the ferromagnet is
stimulated to precess around the $y$ axis, by using a radio frequency
signal, a spin bias polarized along the $y$ direction will be created
in the covered region of the TI film, and electrical current along x-axis will
be generated due to the IEE.
}\label{fig1}
\end{center}
\end{figure}

 Let us start from the effective Hamiltonian of surface states of a thin film 
of 3D TI $\text{Bi}_2\text{Se}_3$~\cite{ShenSQ, LiHC}
\begin{equation}
H = \frac{\Delta}{2} \hat{\tau}_z \hat{\sigma}_z +
     v_{f}\left(p_y \hat{\sigma}_x -p_x \hat{\sigma}_y \right) ~ .
\end{equation}
Here, $\vec{p} = (p_x, p_y)$ is the electron momentum, $\hat{\sigma}_{\alpha}$ with $\alpha=x,y,z$ are the Pauli matrices for electron spin,  and $\hat{\tau}_z$ describes the bonding and antibonding of surface states on the two surfaces,
with $\Delta$ as the hybridization energy.
The eigenenergies for $\tau_z=\pm 1$ are degenerate, given by
\begin{equation}
E_{\tau_z}\left(\vec{p}\right) =
\pm \sqrt{v_{f}^2 p^2 +\frac{\Delta^2}{4}}\ .
\end{equation}
Here, $p^2 = p_x^2+p_y^2$, and signs $+$ and $-$ are for the conduction and valence bands, respectively.
The corresponding eigenstates will be denoted as $\vert\tau_{z}\rangle$. The Fermi energy $E_{\mathrm{F}}$ is set to be in the conduction band. We now calculate the
average of $\hat{\sigma}_{y}$ in the eigenstates $\vert\tau_{z}\rangle$ by using
 the Feynman-Hellman Theorem, yieding $\langle \tau_{z}\vert\hat{\sigma}_y
\vert \tau_{z}\rangle = -v_x/v_{f}$ with $v_{x}=v_{f}^2p_{x}/E_{\mathrm{F}}$, which will be used later.
The Fermi velocity, being renormalized by the nonzero hybridization energy,
becomes $v_{\mathrm{F}} = v_{f}^2p_{\mathrm{F}} /E_{\mathrm{F}}$.

Fig.\ \ref{fig1} illustrates the setup for observing the IEE.
A ferromagnet covers a part of a TI film. When the magnetization
is stimulated to precess around a certain direction $\hat{\bf n}$,
a spin bias polarized along $\hat{\bf n}$
is generated in the covered region of the TI film.
In other words, for an electron with spin parallel or antiparallel
to $\hat{\bf n}$, its chemical potential increases or decreases
by an amount $(-e\mu)$. The spin bias can be conveniently described
by the operator $(-e\mu)\hat{\mbox{\boldmath{$\sigma$}}}\cdot
\hat{\bf n}$. In the ballistic regime, it has been demonstrated
that for the geometry shown in Fig.\ 1, only the $y$ component of the spin bas
contributes to the IEE effect~\cite{Luo}. Therefore, for simplicity,
we focus on the favorable situation, where the spin bias is polarized in the $y$ direction.
The semi-classical boltzmann equation~\cite{Geng} is used to
describe the electronic transport
\begin{equation}
v_x\frac{\partial f_{\tau_z}}{\partial x} =
-\frac{f_{\tau_z}-\bar{f}_{\tau_z}}{\tau_{0}}\ .
\label{Boltz0}
\end{equation}
where $f_{\tau_z}(x,v_{x})$ is the nonequilibrium distribution function of the electrons
in the $\tau_{z}$ band, and $\tau_{0}$ is the relaxation time
 due to impurity scattering. In the linear-response regime,
the distribution function takes the form
$f_{\tau_z} = f_0 + \left(-\frac{\partial f_0}{\partial E_{\tau_{z}}}\right)g_{\tau_z}\left(x,v_x\right)$,
where $f_{0}$ is the equilibrium distribution function.
It follows from Eq.\ (\ref{Boltz0}) that $g_{\tau_z}(x,v_{x})$ satisfies the
following equation
\begin{equation}
v_x\frac{\partial g_{\tau_z}}{\partial x} =
-\frac{g_{\tau_z}-\bar{g}_{\tau_z}}{\tau}\ ,\label{Boltz1}
\end{equation}
where $\bar{g}_{\tau_z}(x)=\frac{1}{2\pi}\int_0^{2\pi} d\phi g_{\tau_z}(x,v_{\mathrm{F}}\cos \phi )$.

The region covered by the ferromagnet is treated as a reservoir, and the uncovered region
is considered as the sample region. Since there is a spin bias in the reservoir, so the electron
distribution function in the reservoir deviates from the equilibrium distribution function,
$f_{\tau_z} \left(x,v_x>0\right) = f_0 +\left(-\frac{\partial f_0}{\partial E_{\tau_{z}}}\right)(-e\mu) \langle
\tau_{z}\vert\hat{\sigma}_y\vert\tau_{z}\rangle$, where the spin bias is projected into the subspace of
the $\tau_{z}$ band. For right-moving electrons, when they just cross
the $x=0$ boundary between the reservoir and sample region,
their distribution function remains to be same as in the reservoir.
As a consequence,
\begin{equation}
g_{\tau_z}\left(x=0,v_x>0\right) = e\mu \frac{v_x}{v_{\mathrm{f}}}\ .
\label{Boundary1}
\end{equation}
The right end of the sample region at  $x=L_{x}$ is assumed to connect to
another equilibrium reservoir. When left-moving electrons cross the boundary $x=L_{x}$, their
distribution function remains to be in the equilibrium state, such that
\begin{equation}
g_{\tau_z} \left(x=L_x,v_x<0\right) = 0\ .
\label{Boundary2}
\end{equation}

Integrating the first-order linear differential equation (\ref{Boltz1}) and taking the boundary conditions
Eqs.\ (\ref{Boundary1}) and (\ref{Boundary2}) into consideration, it is easy to obtain
a self-consistent equation for $\overline{g}_{\tau_{z}}(x)$, which can be solved numerically~\cite{Geng}.
In Ref.\ \cite{Geng}, it is found that a linear approximation $\bar{g}_{\tau_z}(x) = a+ bx$
to $\overline{g}_{\tau_{z}}(x)$
generally works very well. In particular, the linear approximation becomes exact
in the ballistic limit, i.e., $L_{y}\ll l_{f}$, and diffusive limit, $L_{y}\gg l_{f}$~\cite{Geng}.
By following a similar procedure as that detailed
in Ref.\ \cite{Geng}, we obtain for the coefficients $a$ and $b$ as
$a = U_L(L_x+\kappa l_f)(L_x+2\kappa l_f)$ and
$b = -U_L/(L_x+2\kappa l_f) $,
where $l_f = v_{\mathrm{F}} \tau_{0}$ is the electron mean free path, $U_L =
(e\mu) \eta\sqrt{1-{\Delta^2}/{4 E_{\mathrm{F}}^2}}$, and
\begin{equation}
\eta =
\frac{\int_{-\pi/2}^{\pi/2} \cos \phi
e^{-\frac{L_x}{2l_f \cos \phi}} d\phi}
{\int_{-\pi/2}^{\pi/2}
e^{-\frac{L_x}{2l_f \cos \phi} } d\phi}\ ,\label{eta}
\end{equation}
\begin{equation}
\kappa =
\frac{\int_{-\pi/2}^{\pi/2}
e^{-\frac{L_x}{2l_f \cos \phi}} d\phi}
{\int_{-\pi/2}^{\pi/2} \frac{1}{\cos \phi}
e^{-\frac{L_x}{2l_f \cos \phi} } d\phi}\ .\label{kappa}
\end{equation}
In the appendix, we will show that this linear approximation to $\bar{g}_{\tau_z}$ is
very accurate in comparison with the exact solution.

The electrical current is given by
\begin{equation}\label{current}
I = \frac{eL_y}{h^2}\sum_{\tau_z}\int v_x g_{\tau_z}(x,v_x) \left(-\frac{\partial f_0}
{\partial E_{\tau_{z}}}\right)dp_xdp_y\ .
\end{equation}
Following Shen, Vignale, and Raimondi~\cite{Shen}, we define an IEE conductance
$G_{\mathrm{IEE}}=I/\mu$. By using the above linear approximation to $\bar{g}_{\tau_z}$,
analytical expression for $G_{\mathrm{IEE}}$ can be obtained as
\begin{equation}\label{GIEE}
G_{\mathrm{IEE}} = G_{\mathrm{IEE}}^0\left(  \chi^{bal}_{\mathrm{IEE}} + \chi^{dif}_{\mathrm{IEE}} \right)
\end{equation}
where $G_{\mathrm{IEE}}^{0} =  G^{0} \sqrt{1-\Delta^2
/4E_{\mathrm{F}}^2}$ with $G^{0}=N_{\mathrm{ch}}(e^2/h)$ and
$N_{\mathrm{ch}}=4k_{\mathrm{F}}L_{y}/h$ as the number of conducting channels, and
\begin{eqnarray}
  \chi_{\mathrm{IEE}}^{bal} &=& \frac{1}{2}\int_{-\frac{\pi}{2}}^{\frac{\pi}{2}}
              \left(\cos\phi- \frac{\eta L_x}{L_x+2 \kappa l_f} \right)
              e^{-\frac{L_x}{2l_f\cos\phi}}\cos\phi d\phi\ , \nonumber\\
  \chi_{\mathrm{IEE}}^{dif} &=& \frac{\eta l_f}{L_x+2 \kappa l_f}\int_{-\frac{\pi}{2}}^{\frac{\pi}{2}}
              \left(1-e^{-\frac{L_x}{2l_f\cos\phi}}\right)\cos^2\phi d\phi~.\nonumber
\end{eqnarray}
We have divided $G_{\mathrm{IEE}}$ into two parts, labeled by superscripts ``bal'' and ``dif'',
corresponding to contributions from electron ballistic and diffusive transport processes.
In the ballistic limit, where $L_x\ll l_{f}$, it is easy to obtain
$G_{\mathrm{IEE}}=\frac{\pi}{4}  G_{\mathrm{IEE}}^0$. This result
is consistent with that obtained by Luo $et$ $al.$~\cite{Luo}
using the Landauer-B\"{u}ttiker formula
in the ballistic regime in the absence of the contact potential barrier.
In the opposite diffusive limit, where $L_{x}\gg l_f$, we have
$G_{\mathrm{IEE}}=\frac{\pi}{2}\frac{l_{f}}{L_{x}}G_{\mathrm{IEE}}^0$,
which is essentially a Drude like formula.

When the electric current $I$ flows through the system, it causes a voltage difference $V=I/G$
between the two ends of the system, where $G$ is the electrical conductance of the system.
We introduce the ratio $\gamma=V/\mu$ to measure the efficiency of the spin-charge conversion.
In general, $\gamma\le 1$, and
$\gamma=1$ would mean perfect spin-charge conversion, 
in which a spin bias $\mu$ is fully converted to an equal amount of charge bias.
Because $I=\mu G_{\mathrm{IEE}}$ by definition, the efficiency ratio can also be expressed
as $\gamma=G_{\mathrm{IEE}}/G$. The expression for $G$ is given by~\cite{Geng}
\begin{equation}\label{G}
G = G^{0}\left(  \chi^{bal} + \chi^{dif} \right)
\end{equation}
where
\begin{eqnarray}
  \chi^{bal} &=& \frac{\kappa l_f}{L_x+2 \kappa l_f}\int_{-\frac{\pi}{2}}^{\frac{\pi}{2}}
              e^{-\frac{L_x}{2l_f\cos\phi}}\cos\phi d\phi\ ,\nonumber \\
  \chi^{dif} &=& \frac{ l_f}{L_x+2 \kappa l_f}\int_{-\frac{\pi}{2}}^{\frac{\pi}{2}}
              \left(1-e^{-\frac{L_x}{2l_f\cos\phi}}\right)\cos^2\phi d\phi~.\nonumber
\end{eqnarray}
Using Eqs.\ (\ref{GIEE}) and (\ref{G}), one can calculate the efficiency ratio.
It is easy to find that the efficiency ratio $\gamma$ normalized by $\gamma_{0}=\sqrt{1-\Delta^2
/4E_{\mathrm{F}}^2}$ is a universal function of $l_{f}/L_{x}$, independent of any model parameters.
The calculated curve of the universal function is displayed in Fig.\ 2. We see that in the ballistic
and diffusive limits, $\gamma/\gamma_{0}$ converges to two different constants. In fact, using
the expressions for $G$ in the two limits~\cite{Geng}, $G=G_{0}$ for $L_{x}\ll l_{f}$, and
$G=\frac{\pi}{2}\frac{l_{f}}{L_{x}}G^0$ for $L_{x}\gg l_{f}$,
 one can readily obtain $\gamma/\gamma_{0}=\frac{\pi}{4}$ in the ballistic limit, and $\gamma/\gamma_{0}=1$
in the diffusive limit. We mention that these asymptotic formulas for $\gamma/\gamma_{0}$ are exact,
because the linear approximation to $\bar{g}_{\tau_z}$ becomes exact in the ballistic and
diffusive limits~\cite{Geng}. The result that $\gamma$ approaches $\gamma_{0}$ in the diffusive limit
can be understood as follows. In the diffusive limit,   $L_{x}\gg l_{f}$, the electrons propagating at small
angles with the $x$ axis, i.e., $\phi\simeq 0$, make dominant contributions to the electric current.
For $\phi\simeq 0$, the boundary condition Eq.\ (\ref{Boundary1}) reduces to
$g_{\tau_z}\left(x=0,v_x>0\right) = e\mu(v_{\mathrm{F}}/v_{f})\cos\phi
\simeq e\mu(v_{\mathrm{F}}/v_{f})= e\mu\gamma_{0}$. Therefore, the spin bias $\mu$ is just equivalent
to a charge bias $\gamma_{0}\mu$, and as a result, the efficiency ratio
becomes $\gamma=\gamma_{0}\mu/\mu=\gamma_{0}$.
When the electron Fermi energy $E_{\mathrm{F}}$ is much larger than
the hybridization gap $\Delta$, we have $\gamma_{0}=1$, so that $\gamma=\frac{\pi}{4}$ in the
ballistic limit and $\gamma=1$ in the diffusive limit. The spin-charge conversion
is perfect in the diffusive limit.

\begin{figure}
\begin{center}
\includegraphics[width=7.0cm]{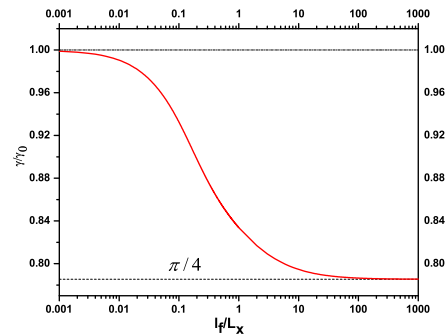}
\caption{The universal function of $\gamma/\gamma_{0}$ versus $l_{f}/L_{x}$, where $ \gamma_0 = \sqrt{1-\Delta^2/4E_f^2} $.
$\gamma/\gamma_{0}$ approaches $\pi/4$ in the ballistic limit, and $1$ in the diffusive limit.
}\label{gamma}
\end{center}
\end{figure}

We have shown that highly efficient IEE or spin-charge conversion can be achieved on a TI surface because of the spin-momentum interlocking of the surface states. An analytical theory for the IEE is developed, which
is valid from the ballistic to diffusive regime. The IEE will be very useful for electrical detection
of spin current and spin accumulation in spintronics.

This work was supported by the State Key Program for Basic
Researches of China under grants numbers 2015CB921202 and
2014CB921103 (L.S.), the National Natural Science Foundation of
China under grant numbers 11674160 (L.S.) and 11474149 (R.S.),
and a project funded by the PAPD of Jiangsu Higher Education
Institutions (L.S. and D.Y.X.).

\appendix
\section{Verification of Linear Approximation with Exact Solution}\label{appendix}
\begin{figure}
\begin{center}
\includegraphics[width=7.0cm]{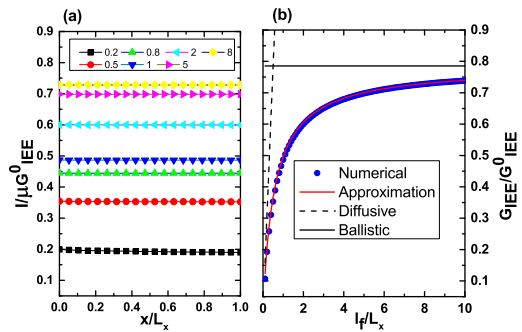}
\caption{(a) Normalized electrical current due to the IEE as a function of normalized
coordinate $x/L_{x}$ for several different values of $l_{f}/L_{x}$. (b) IEE conductances as
functions of $l_{f}/L_{x}$ calculated from exact numerical solution and approximate formula
Eq.\ (\ref{GIEE}). The black solid line stands for the result of the Landauer-B\"{u}ttiker like formula
in the ballistic regime, $G_{\mathrm{IEE}}/G_{\mathrm{IEE}}^{0}=\frac{\pi}{4}$, and the
black dash line stands for the Drude like formula in the diffusive regime,
$G_{\mathrm{IEE}}/G_{\mathrm{IEE}}^{0}=\frac{\pi l_{f}}{2L_{x}}$.
}\label{current_conservation}
\end{center}
\end{figure}

In this appendix, we show that our linear approximation $\bar{g}_{\tau_z} = a+bx$ 
is a very good approximation compared with the exact numerical result. 
In Fig.\ \ref{current_conservation}(a), we show the
exactly calculated electrical current $I(x)$ due to the IEE as
a function of position $x$, for several different values of $l_{f}/L_{x}$.
For a given value of $l_{f}/L_{x}$, $I(x)$ is a constant independent of $x$, meaning that
the continuity of the electrical current is satisfied. This serves as an evidence that our
numerical result is accurate. In Fig.\ \ref{current_conservation}(b), we plot
$G_{\mathrm{IEE}}/G^{0}_{\mathrm{IEE}}$ calculated from the exact solution and
approximate formula Eq.\ (\ref{GIEE}) as functions of $l_{f}/L_{x}$.
The approximate formula Eq.\ (\ref{GIEE}) fits very well with
the exact solution.

\begin{figure}
\begin{center}
\includegraphics[width=7.0cm]{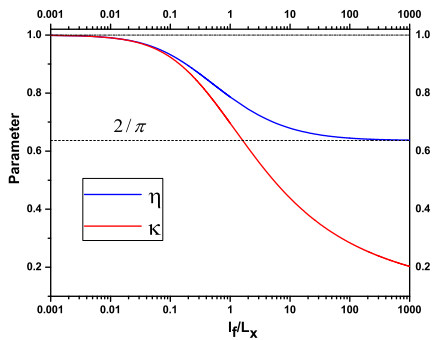}
\caption{ Parameters $\eta$ and $\kappa$ as functions of $l_f/L_x$.
}\label{paras}
\end{center}
\end{figure}

Finally, we plot the curves for the two parameters $\eta$ and $\kappa$ given in Eqs.\ (\ref{eta}) and (\ref{kappa})
in Fig.\ \ref{paras} for reference.
We can see that in the ballistic limit $L_{x}\ll l_f$,
$\eta \rightarrow 2/\pi$ and $\kappa \rightarrow 0$. In the diffusive limit $L_{x}\gg l_f$, $\eta \rightarrow 1 $ and $\kappa \rightarrow 1$. These results can also be derived
directly from the expressions Eqs.\ (\ref{eta}) and (\ref{kappa}).

\end{document}